\documentclass[aps,twocolumn]{revtex4-2}
\usepackage{graphicx}
\usepackage[justification=justified,width=\linewidth]{caption}
\usepackage{subcaption}
\usepackage{amsmath}
\usepackage{amsfonts}
\usepackage{amsthm}
\usepackage{amssymb}
\usepackage{amsbsy}
\usepackage{wasysym}
\usepackage{bm}
\usepackage{mathrsfs}
\usepackage{color}
\usepackage{times}
\usepackage{hyperref}

\begin{document}

\title{Breaking the Logarithmic Barrier: Activity-Induced Recovery of Phase Separation Dynamics in Confined Geometry}
\author{Preethi M, Parameshwaran A and Bhaskar Sen Gupta}
\email{bhaskar.sengupta@vit.ac.in}
\affiliation{Department of Physics, School of Advanced Sciences, Vellore Institute of Technology, Vellore, Tamil Nadu - 632014, India}

\begin{abstract}
Phase separation in confined environments is a fundamental process underlying geological flows, porous filtration, emulsions, and intracellular organization. Yet, how confinement and activity jointly govern coarsening kinetics and interfacial morphology remains poorly understood. Here, we use large-scale molecular dynamics simulations to investigate vapor-liquid phase separation of passive and active fluids embedded in complex porous media. By generating porous host structures via a freeze-quench protocol, we systematically control the average pore size and demonstrate that confinement induces a crossover from the Lifshitz-Slyozov power-law growth to logarithmically slowed coarsening, ultimately arresting domain evolution. Analysis of correlation functions and structure factors reveals that confined passive systems exhibit fractal interfaces, violating Porod’s law and indicating rough morphological arrest. In contrast, introducing self-propulsion dramatically changes the coarsening pathway: activity restores smooth interfaces, breaks the confinement-induced scaling laws, and drives a transition from logarithmic to ballistic domain growth at high activity levels. Our findings reveal an activity-controlled mechanism to overcome geometric restrictions and unlock coarsening in structurally heterogeneous environments. These insights establish a unifying framework for nonequilibrium phase transitions in porous settings, with broad relevance to active colloids, catalytic media, and biologically crowded systems, where living matter routinely reorganizes within geometric constraints to sustain function.
\end{abstract}

\maketitle

\section{Introduction}
Phase separation is a ubiquitous nonequilibrium process that governs structure formation across a diverse range of natural and engineered systems, from industrial emulsions~\cite{Emulsion} and geological flows~\cite{Geological-flow} to intracellular biomolecular condensates~\cite{biomolecular-condensate}. When a homogeneous fluid is quenched below its critical temperature, vapor-liquid coexistence emerges through nucleation and growth or spinodal decomposition, followed by coarsening of the domains~\cite{Majumder,Roy-VL}. In bulk fluids, the characteristic domain length scale $\ell(t)$ exhibits well-established growth laws: diffusive kinetics at early times, $\ell(t) \sim t^{1/3}$~\cite{Lifshitz}, crosses over to viscous hydrodynamic growth, $\ell(t) \sim t$~\cite{Majumder}, as momentum transport accelerates coarsening. These universal power-law scalings brace our understanding of equilibrium demixing in unbounded media.

However, most real systems are embedded within heterogeneous and tortuous geometries. In porous materials, geological reservoirs, functional filters, and crowded intracellular environments, confinement restricts transport pathways, screens hydrodynamics, and introduces geometric bottlenecks that impede domain coalescence~\cite{H.Tanaka,Bhattacharya1,Davis1}. Prior studies have demonstrated that such spatial disorder can arrest the coarsening of vapor–liquid phase separation, induce metastable morphologies, and lead to violations of classical scaling laws~\cite{preethi}.

Biological systems raise this challenge even further. Active fluids, composed of self-propelled or force-generating constituents continuously consume energy at microscopic scales, profoundly altering their collective organization~\cite{Cates2015,Marchetti,Ramaswamy,paramesh}. Examples span many length scales: motile bacteria confined in microfluidic chambers form dynamic clusters guided by geometry~\cite{microfluidic-chamber}; active colloids at interfaces produce structured aggregates relevant to materials design; and inside living cells~\cite{active-colliods}, membraneless organelles (e.g., nucleoli, stress granules) exhibit features reminiscent of active vapor-liquid phase separation, where motor proteins, cytoskeletal forces, and biochemical activity sustain growth, mobility, and turnover of condensates. These systems demonstrate how activity, confinement, and phase separation synergistically shape biological organization. The extent to which activity may enable coarsening to overcome geometric restrictions, effectively “unlocking” arrested phase separation in confined media, remains an open question with broad physical and biological implications.

Active matter is often modeled using one of three microscopic driving mechanisms: (1) Active Brownian particles, where propulsion is governed by rotational noise~\cite{Cates2013-ABP}; (2) Run-and-tumble dynamics, relevant for bacterial motility~\cite{Cates2008-RTP}; and (3) Vicsek-type alignment, where particles self-propel while aligning their motion with neighbors~\cite{vicsek1995}. The Vicsek model plays a unique role in connecting active phase separation to biological systems. It captures collective alignment, a hallmark of flocking bacteria, cytoskeletal filaments interacting through motors, and cells undergoing coordinated migration. It also enables momentum-like transfer through clusters and enhanced interfacial persistence, features increasingly recognized in active biomolecular condensates and cytoskeletal assemblies. Thus, studying vapor-liquid phase separation in a Vicsek-driven fluid offers a mechanistic framework for understanding how biological activity can reshape phase separation kinetics under confinement.

Additionally, understanding how confinement and activity jointly regulate phase separation is critical for interpreting processes ranging from droplet coalescence in porous filtration and enhanced oil recovery~\cite{oil-recovery}, to intracellular condensate dynamics in biological tissues where geometry and motility are inherently coupled. Moreover, the ability to tune coarsening kinetics through environmental structure or particle activity could inform the design of adaptive materials, controlled release platforms, and microfluidic systems that exploit nonequilibrium pattern formation.

In this work, we perform large-scale molecular dynamics simulations of a model vapor-liquid system embedded in complex porous architectures produced through a freeze-quench protocol. We first quantify how confinement alone transforms passive coarsening, replacing classical diffusive growth with a logarithmic regime reflecting kinetic arrest. Introducing Vicsek-type activity uncovers dramatic reversal: active alignment sharpens interfaces, restores fast power-law growth, and ultimately allows domains to overcome geometric constrictions imposed by pores. Together, these results establish a unified picture of how activity can unlock confined phase separation, providing insight into diverse living and engineered systems where geometry and nonequilibrium driving are inherently coupled.

\section{Model and method}
\subsection{Model for Activity} To model self-driven behavior in the confined vapor-liquid system, we employ a one-component active fluid in which particles align their direction of motion with neighbors, inspired by the Vicsek alignment model~\cite{vicsek1995,czirok2000}. In addition to this active alignment, particles interact via a passive Lennard–Jones (LJ) potential, enabling the system to exhibit both vapor-liquid coexistence and nonequilibrium collective motion~\cite{paramesh,Das2017}. Such a combination is often used to capture clustering phenomena in active matter, including motile colloids, swarming cells, or self-propelled nanoscopic particles, where alignment and interparticle attractions jointly influence emergent structure.

Self-propulsion is introduced through a Vicsek-type force $\Vec{F}^S=f_A \hat{v}_{r_c}$, where $f_A$ denotes the activity strength and $\hat{v}_{r_c}$ is the normalized average velocity of neighboring particles within an interaction range $r_c$:
\vspace*{-6pt}
\begin{equation}
	\hat{v}_{r_c} = \frac{\sum_j \Vec{v}_j} {|\sum_j \Vec{v}_j} | \
\end{equation}
Naively adding this term to the velocity update,
\begin{equation} \label{eq2}
	\Vec{v_i}(t+\Delta t)= \Vec{v_i}^{pas}(t+\Delta t) +\frac{\Vec{F_i}^S}{m_i}\Delta t 
\end{equation}
would not only reorient the velocity but also increase its magnitude. This leads to an uncontrolled rise in kinetic temperature, thereby affecting the phase behavior. To avoid this artifact and retain the essential Vicsek physics, we constrain the particle speed to its passive value~\cite{paramesh,Das2017}:
\begin{equation}
	\Vec{v_i}(t+\Delta t)=|\Vec{v_i}^{pas}(t+\Delta t)| \hat{n} ,
\end{equation}
where
\begin{equation}
	\hat{n}=\Bigl(\Vec{v_i}^{pas}(t+\Delta t) +\frac{\Vec{F_i}^S}{m_i}\Delta t\Bigl) \Bigl/ \Bigl( |\Vec{v_i}^{pas}(t+\Delta t) +\frac{\Vec{F_i}^S}{m_i}\Delta t|\Bigl)
\end{equation}
Thus, by following this method, the self-propelling force applied to each particle alters only the direction of motion, similar to the original Vicsek model. This approach enables us to control the system temperature more accurately and study the phase behavior.

This prescription ensures that activity modifies only the orientation of particle motion, preserving temperature control and allowing clean exploration of nonequilibrium effects. To maintain the system at a prescribed temperature $T$, we employ a Langevin thermostat~\cite{allen1987,frenkel2002}, integrating the particle dynamics via
\begin{equation} \label{eq:1}
	\dot{\Vec{v}}_i = -\frac{\Vec{\nabla} V_i}{m_i}- \zeta\Vec{v}_i + \sqrt{\frac{2\zeta K_B T}{m_i}} \, \Vec{\xi}_i(t) + \Vec{F}^S_i  \bm{;} \hspace{0.2cm}  \Vec{v}_i = \dot{\Vec{r}}_i  \hspace{0.2cm}
\end{equation}
Here $m_i$ is the mass of the $i^{th}$ particle, $V_i$ is the LJ interaction potential, $\zeta$ is the damping coefficient, and $\Vec{\xi}$ is a delta-correlated Gaussian noise satisfying
\begin{equation}
\langle \Vec{\xi} _{i\mu}(t)\Vec{\xi} _{j\nu}(t') \rangle =\delta_{\mu\nu}\delta_{ij}\delta (t-t')
\end{equation}
Thus, the model incorporates alignment-driven activity while rigorously maintaining thermal control, a key requirement for studying phase separation in active fluids.

\subsection{Creation of Porous Structures} To systematically investigate the role of geometric confinement in phase separation kinetics, we construct porous environments through a self-assembly followed by a freeze-quench protocol~\cite{Bhattacharya1}. We begin with a binary mixture of equal-sized particles, species $A$ and $B$, in a 50:50 composition at number density $\rho=N/V=1$. The pairwise interaction is governed by the standard LJ potential,
\begin{equation}
	U_{\alpha\beta}(r) = 4\epsilon_{\alpha\beta} \left[\left(\frac{\sigma_{\alpha\beta}}{r}\right)^{12} - \left(\frac{\sigma_{\alpha\beta}}{r}\right)^6\right]
	\label{1}
\end{equation}
where $\sigma_{\alpha\beta}$ and $\epsilon_{\alpha\beta}$ represent the particle diameter and interaction strength, respectively. The indices $\alpha, \beta \in \{A, B\}$, and $r = |\vec{r}_i - \vec{r}_j|$ represents the separation distance between two particles. To promote phase separation, we set $\epsilon_{AA} = \epsilon_{BB} = 1.0$, and $\epsilon_{AB} = 0.5$, and $\sigma_{AA} = \sigma_{BB} = \sigma_{AB} = 1.0$, corresponding to a critical temperature $T_c = 1.42$ (binary liquid)~\cite{Tc-LL}. A cutoff $r_c$ = 2.5$\sigma$ is used for computational efficiency. Throughout, we measure length in units of $\sigma$ and temperature in $\epsilon/k_B$,with $k_B$ signifying Boltzmann’s constant. For convenience, the mass of each particle, $\epsilon$, $\sigma$, and $k_B$ are set to unity.

The simulation begins by preparing a uniform system of $N = 262,144$ passive particles, organized within a cubic simulation box of size $L = 64$ in the canonical ensemble. Molecular dynamics (MD) simulations are employed to equilibrate the binary mixture at a high temperature of $T=10.0$, where the system behaves as a homogeneous fluid. After equilibration, the system is quenched at a subcritical temperature $T=0.77T_c$, initiating spontaneous phase separation into interconnected $A$-rich and $B$-rich domains. Once a well-developed bicontinuous structure is formed, a snapshot of the configuration is selected at a chosen time $\tau$ (in units of $(m\sigma^2/\epsilon)^{1/2}$). The $A$-type particles is then frozen in their positions, effectively creating a static porous matrix. These frozen particles serve as the solid scaffold that defines the confinement geometry for subsequent simulations. The resulting structure exhibits a tortuous network of voids and channels. The average pore size $d_p$ is tuned via the time $\tau$ before freezing. The larger $\tau$ yields larger average pore sizes.

The other species ($B$-type) is manipulated to create the fluid phase. The appropriate number of randomly chosen $B$-type particles is removed from the simulation box to achieve a target density $\rho = 0.3$ of the same. The described fluid system has a bulk critical temperature of $T_c = 0.94\epsilon/k_B$ (vapor-liquid) and a critical density $\rho_c = 0.32$ associated with the vapor-liquid transition in the passive limit~\cite{Roy-VL,Daniya-Gravity}. The interaction strength between the stationary matrix and the fluid is set to $\epsilon_{AB} = 0.5$, with a cutoff distance $r_c = 2^{1/6}\sigma$, effectively removing any attractive forces between them. This ensures that the confined system maintains a well-defined vapor-liquid equilibrium state. In subsequent simulations, these $B$-type particles are endowed with activity according to the Vicsek alignment rule described above.

\subsection{Molecular Dynamics Simulations} The setup mentioned above ensures that the active $B$ particles can undergo vapor-liquid phase separation under confinement. Active fluid dynamics is simulated using MD simulations with the velocity-Verlet algorithm~\cite{Verlet} and a timestep $\Delta t = 0.001$. To initiate phase separation, the $B$-type particles is first reheated to a high temperature of $T=10.0$ to eliminate the memory effects of the freezing step and then quenched to a target temperature $T=0.8$ ($<T_c$). Thermal fluctuations are modeled using a Langevin thermostat, which provides stochastic damping and noise consistent with the desired temperature (see Eq. \ref{eq:1}). Periodic boundary conditions are applied in all directions to minimize finite-size effects. To ensure statistical reliability, we perform ensemble averaging over 30 independent runs starting from uncorrelated initial conditions.

\subsection{Domain morphology and growth dynamics} To study the phase ordering dynamics, we resort to the length scale $\ell(t)$ representing the average size of the domains. This is computed from the first zero crossing of the two-point equal-time correlation function~\cite{Puri-book,ParameshwarA}, given by
\begin{equation}
	C_{\psi\psi}(\vec{r},t) = \langle\psi(0,t)\psi(\vec{r},t)\rangle - \langle\psi(0,t)\rangle \langle\psi(\vec{r},t)\rangle
	\label{corr}
\end{equation} 
where $\psi(\vec{r},t)$ is the order parameter related to the local density at position $\vec{r}$. The angular brackets represent statistical averaging. 

To analyze interface morphology, static structure factor $S(\vec{k},t)$ is calculated by taking the Fourier transform of the correlation function given by~\cite{Bray,Parameshwaran2},
\begin{equation}
	S(\vec{k},t) = \int d\vec{r} \hspace{0.1cm} exp(i\vec{k}.\vec{r}) \hspace{0.1cm} C_{\psi\psi}(\vec{r},t).
	\label{struc}
\end{equation}
For a sharp interface, the structure factor follows the so-called Porod's law in $d$-dimension, characterized by $S(k,t ) \sim k^{-(d+1)}$ in the large k limit~\cite{Bray}. Any deviation from Porod's law indicates roughening of domain boundaries and fractality. 

\section{Results}
\begin{figure}[h]
	\centering
	 \includegraphics[width=26mm]{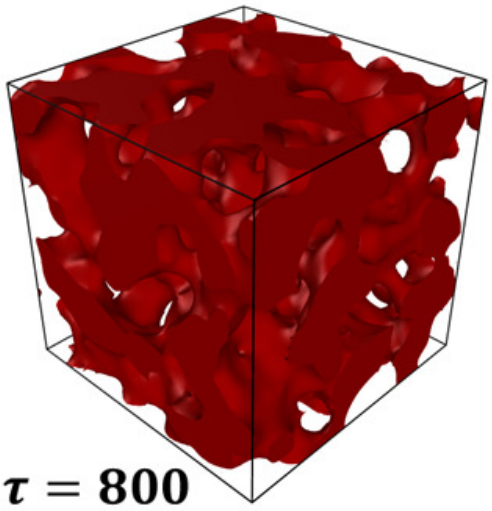}
	  \includegraphics[width=26mm]{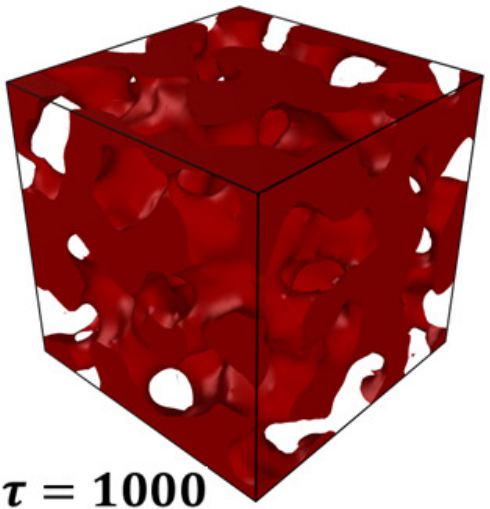}
	   \includegraphics[width=26mm]{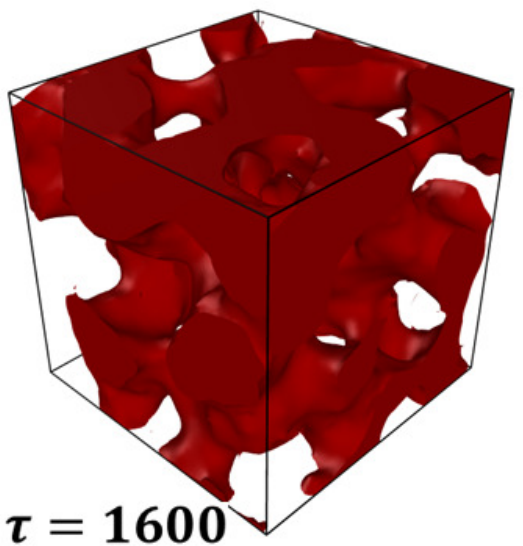}\\
	    \includegraphics[width=26mm]{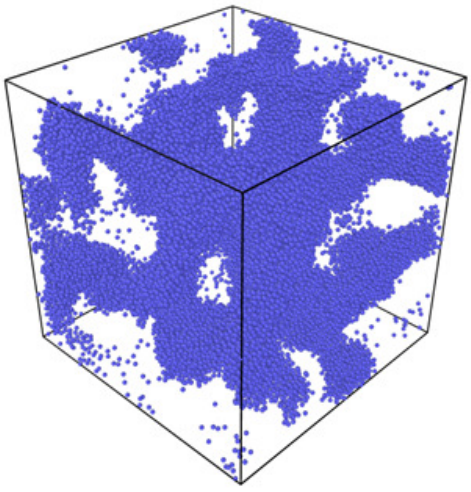}
	     \includegraphics[width=26mm]{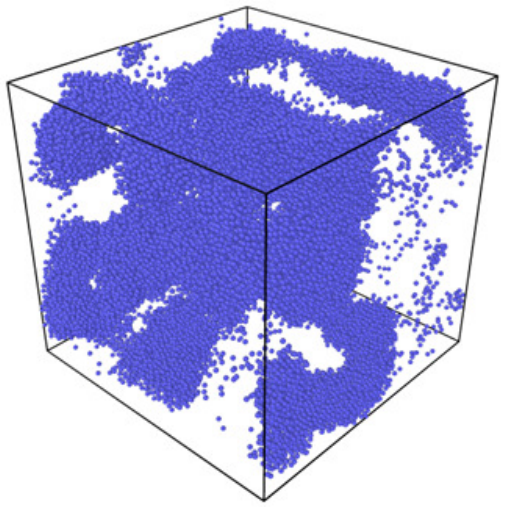}
	      \includegraphics[width=26mm]{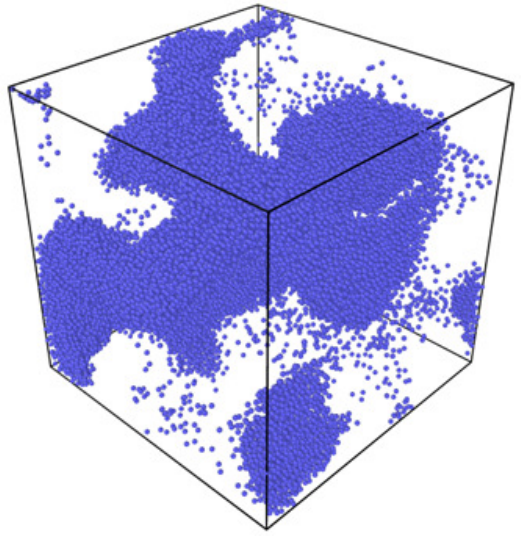}
	\caption{ Illustrates three distinct porous host structures corresponding to different $\tau$ values mentioned in the figure (top panel) and snapshots of the vapor-liquid domains (bottom panel) at time $t$ = 500. }
	\label{fig:1}
\end{figure}

The simulation protocol establishes a direct relationship between the waiting time $\tau$ and the characteristic pore size $d_P$, where the later increases systematically with $\tau$. In this study, we examine three porous host structures generated using $\tau = 800, 1000,$ and $1600$ as shown in Fig.~\ref{fig:1}. The resulting configurations clearly display an increase in average pore size with increasing $\tau$. These structural differences have a pronounced influence on the subsequent vapor–liquid phase separation: at a fixed evolution time $t = 500$, the liquid domains are visibly larger in hosts with larger $d_P$, indicating that the coarsening kinetics is strongly controlled by the underlying pore geometry.

\begin{figure}[t]
	\centering
	\includegraphics[width=40mm,height=30mm]{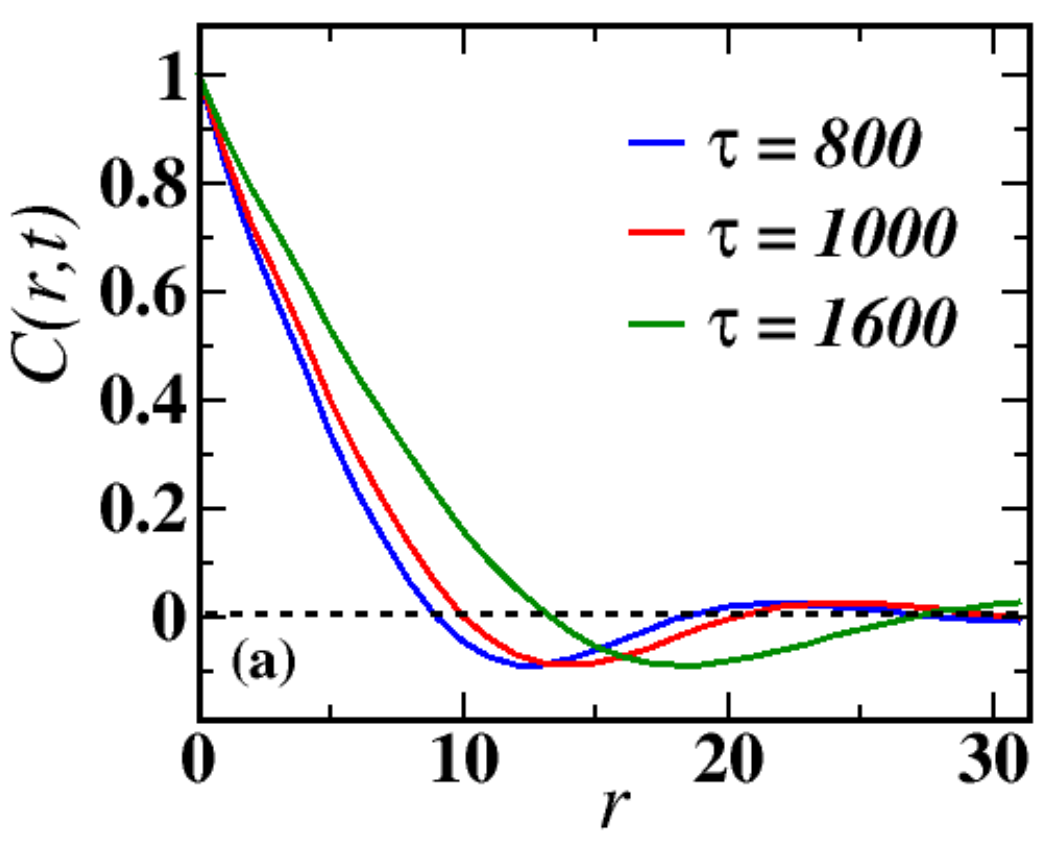}~~~
	\includegraphics[width=38mm,height=29.5mm]{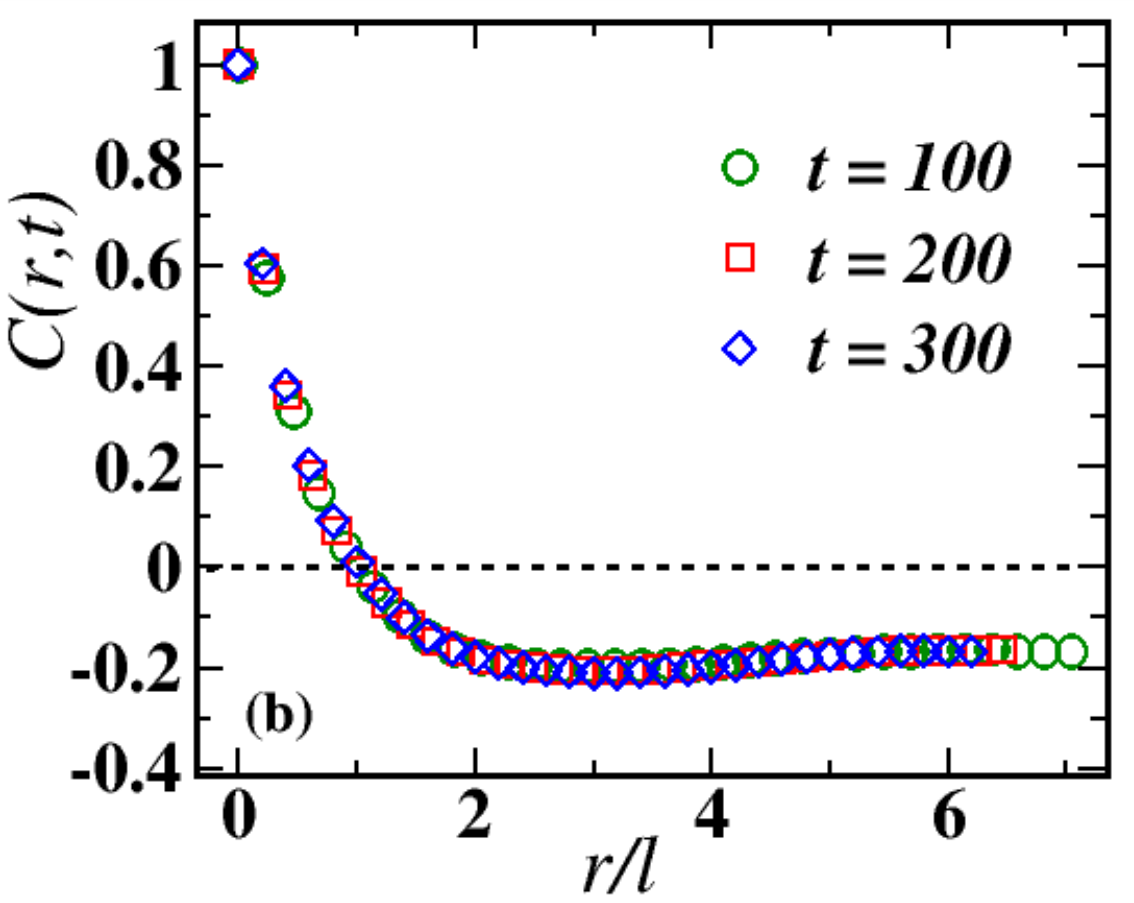}
	\caption{(a) Correlation function $C(r,t)$ vs. $r$ for the frozen $A$-type particles comprising three distinct porous host structure. (b) The scaling plot of $C(r,t)$ vs. $r/\ell$ for the fluid system confined within the porous material with $d_P$ = 9.0 at different time.}
	\label{fig:2}
\end{figure}

To quantify the characteristic length scales, we evaluate the two-point equal-time correlation function $C(r,t)$, defined in Eq.~(\ref{corr}). As a first step, we compute the average pore size of the host structures shown in Fig.~\ref{fig:1}, treating the frozen $A$-type particles as a static domain pattern. The order parameter field $\psi(\mathbf{r},t)$ is constructed from the local density contrast between species A and B within a cubic volume of side $2\sigma$ centered at position $\mathbf{r}$. The field is assigned $\psi = +1$ when species A dominates locally and $\psi = -1$ otherwise. Figure~\ref{fig:2}(a) shows $C(r,t)$ for the $A$-type particles for $\tau = 800, 1000,$ and $1600$. Following standard practice~\cite{Puri-book,Bray}, we extract the average domain size $\ell(t)$ from the first zero crossing of $C(r,t)$. For the frozen host structures, this measure yields pore sizes $d_P \approx \ell(t) \simeq 8.0, 9.0,$ and $12.0$ for $\tau = 800, 1000,$ and $1600$, respectively.

\subsection{Passive Case}
We first examine the coarsening dynamics of the passive vapor–liquid system by setting the activity to zero ($f_A = 0$). The order parameter field $\psi(\mathbf{r},t)$ is constructed to distinguish the liquid-rich and vapor-rich regions as~\cite{Majumder,Roy-VL,Das2017,Daniya-Gravity},
\begin{equation}
	\psi(\vec{r}, t) =
	\begin{cases} 
		0 & \text{wall particles},\\
		+1 & \text{if } \rho(\vec{r}, t) > \rho_c,  \\
		-1 & \text{otherwise}
	\end{cases}
	\label{}
\end{equation}
where $\rho(\vec{r},t)$ is the local density computed within a cubic cell of side $2\sigma$ centered at $\vec{r}$. This coarse-grained representation enables us to monitor the time evolution of domain morphology.
\begin{figure}[t]
	\centering
	\includegraphics[width=38mm,height=30mm]{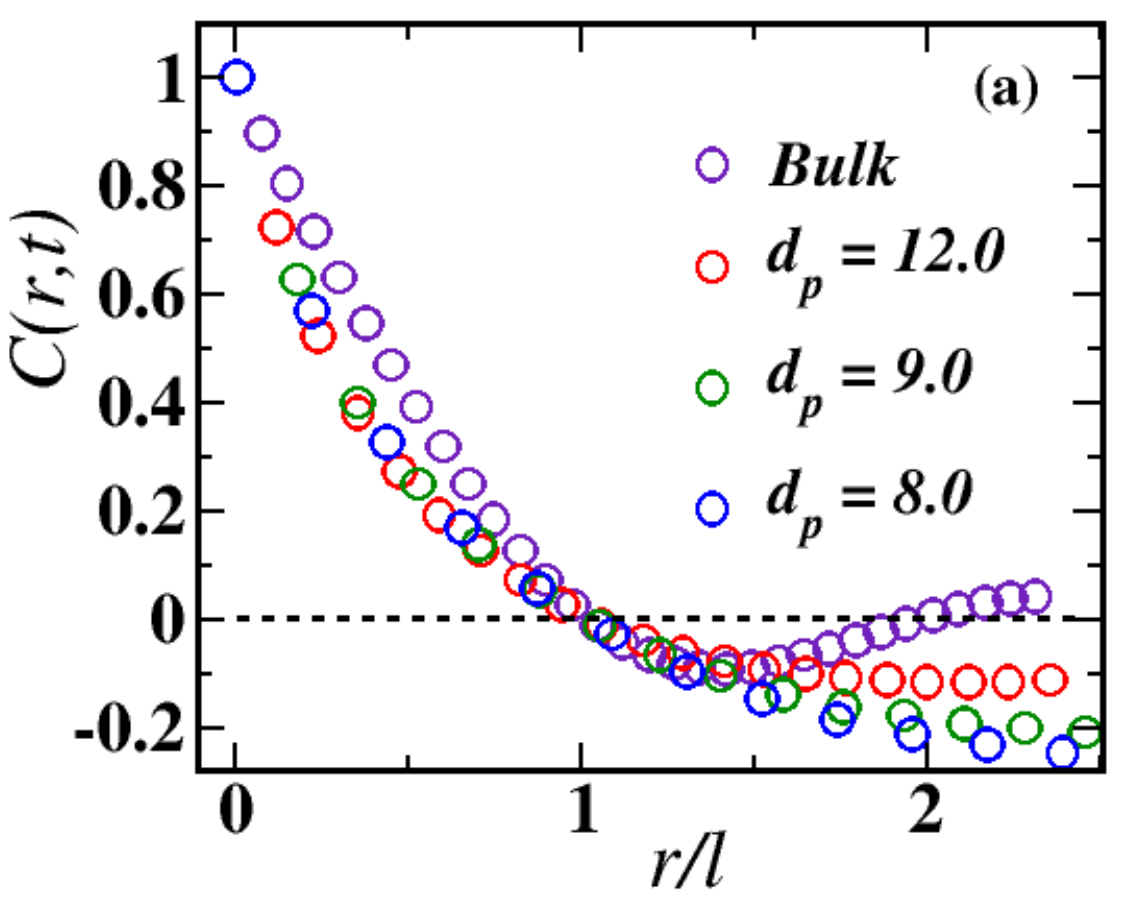}~~~
	{\includegraphics[width=39mm,height=30mm]{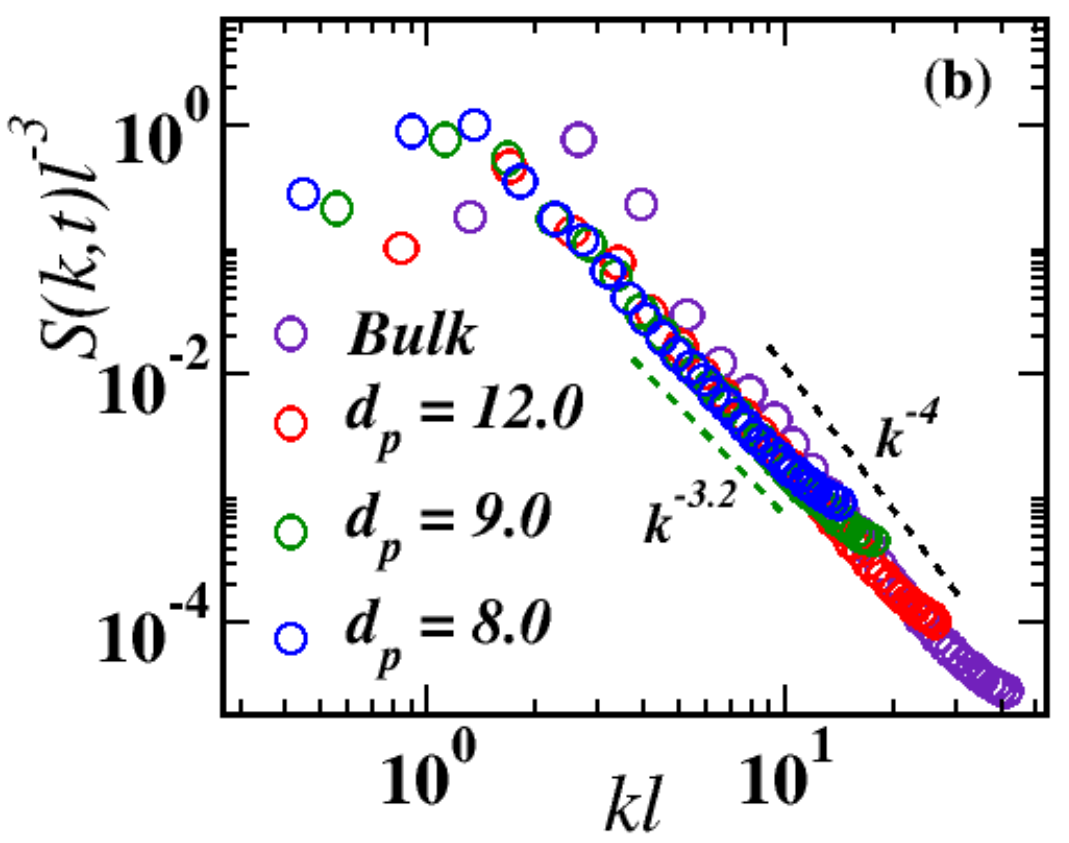}}
	\caption{ (a) The same scaling plot in Fig. 2(b) for the bulk and confined fluid systems at time $t=500$. (b) The scaled structure factor $S(k,t)\ell^{-3}$ vs $kl$ for the system in (a) at time $t$ = 500.}
	\label{fig:3}
\end{figure} 

 Figure~\ref{fig:2}(b) shows the scaling behavior of $C(r,t)$ plotted against $r/\ell(t)$ for a representative pore size, $d_P = 9.0$. The collapse of data over multiple times demonstrates dynamical scaling and confirms that coarsening proceeds self-similarly despite the presence of geometric confinement~\cite{Binder-book}. Comparable scaling behavior is also observed for $d_P = 8.0$ and $12.0$ (not shown), implying that the confined passive fluid retains the same dynamical universality class with respect to scaling morphology.

We next compare bulk and confined systems. Figure~\ref{fig:3}(a) reveals that the data collapse is inadequate when plotted as $C(r,t)$ vs. $r/\ell(t)$, indicating that superuniversality is violated in confined geometries~\cite{CorberiSU1,CorberiSU2}. Consistent with Porod’s law, the bulk system exhibits a linear decay at short distances, a hallmark of scattering from sharp, smooth interfaces~\cite{Bray}. In contrast, confined systems show a pronounced cusp at small $r$, signaling a roughened, fractal interface~\cite{Gaurav,Bhattacharyya2,Bhattacharyya3}.

Complementary insight is provided by the static structure factor $S(k,t)$, shown in Fig.~\ref{fig:3}(b). While bulk behavior follows the expected Porod tail, $S(k,t)\sim k^{-(d+1)}$ at large $k$~\cite{Puri-book}, confinement leads to a non-Porod power law $S(k,t)\sim k^{-(d+\theta)}$ with $\theta\approx0.2$ for $d_P = 9.0$, corresponding to a fractal dimension $d_f = d - \theta \approx 2.8$. This deviation further demonstrates the strong structural influence imposed by the porous host.
\begin{figure}[t]
	\centering
	{\includegraphics[width=36mm,height=30mm]{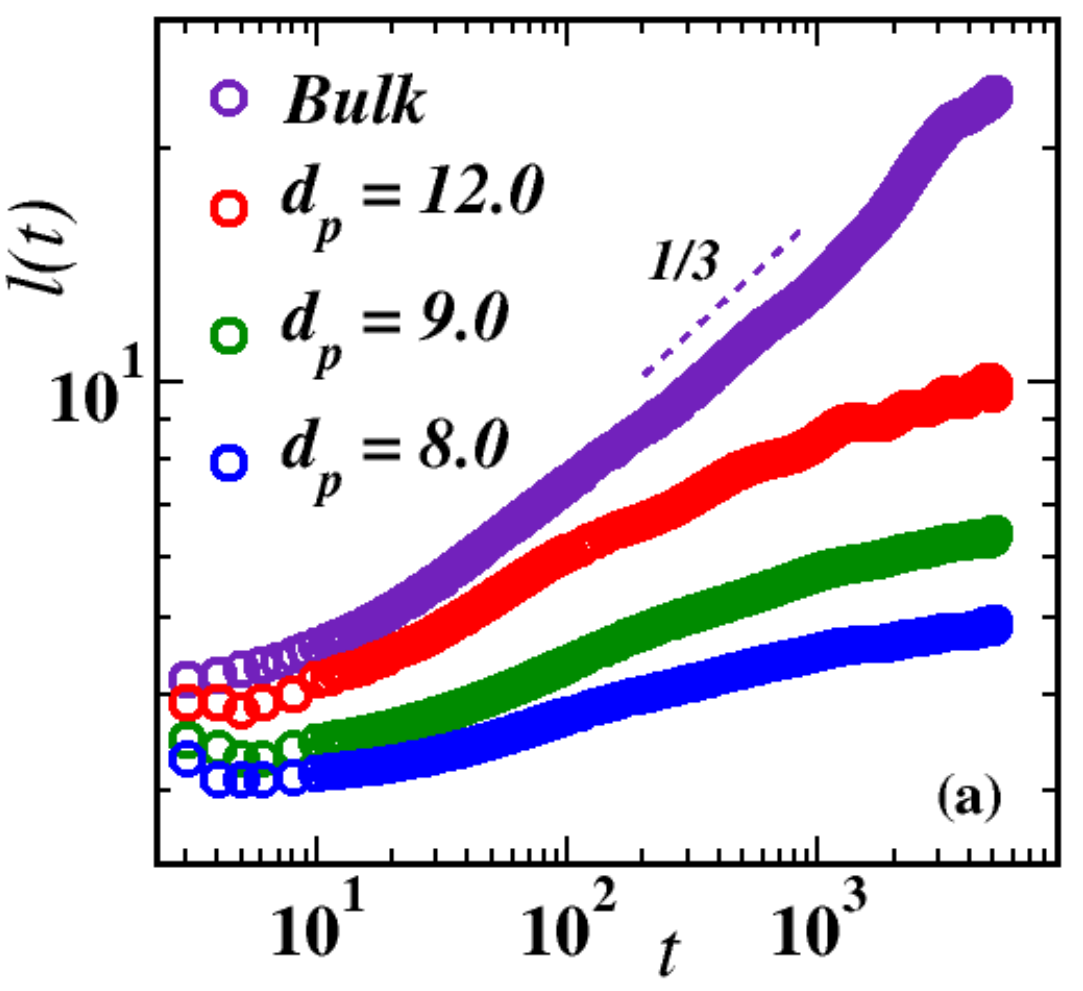}}~~~
		\includegraphics[width=37mm,height=30mm]{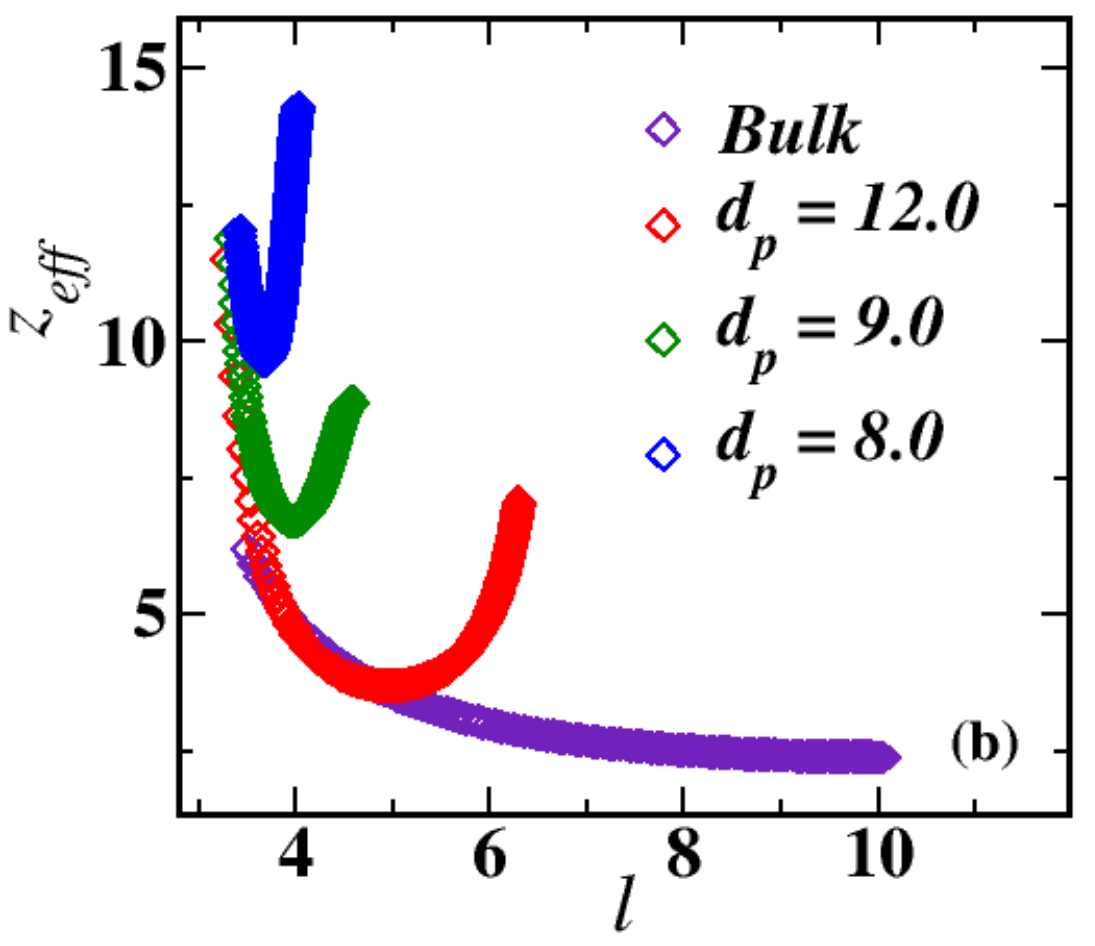}
	\caption{ (a) The time evolution of the average domain size $\ell(t)$ for the bulk and confined system, depicted on a log-log scale. (b) Effective exponent $z_{\mathrm{eff}}$ vs. $\ell$ for the length scale shown in (a).}
	\label{fig:4a}
	\label{fig:4b}
\end{figure} 

To characterize the ordering kinetics, we focus on the characteristic domain length scale $\ell(t)$, extracted from the two-point equal-time correlation function $C(r,t)$. Figure~\ref{fig:4a}(a) displays the time evolution of $\ell(t)$. In bulk, under the influence of the Langevin thermostat, domain growth follows the Lifshitz-Slyozov (LS) diffusion mechanism, characterized by a power-law growth with an exponent of $\alpha=1/3$~\cite{Lifshitz}. Under confinement, the coarsening dynamics become significantly slower. The more restrictive the pore geometry (smaller $d_P$), the more strongly the diffusion is hindered. At long times, $\ell(t)$ saturates at a pore size dependent value, reflecting the suppression of full phase separation and the emergence of metastable connectivity. Physically, confined topologies introduce energy barriers that limit coarsening. Consequently, a modified domain growth law is anticipated for the confined system. The dependence of these barriers on the domain size $\ell(t)$ ultimately determines the asymptotic growth law.
\begin{figure}[t]
	\centering
		\includegraphics[width=28mm,height=25mm]{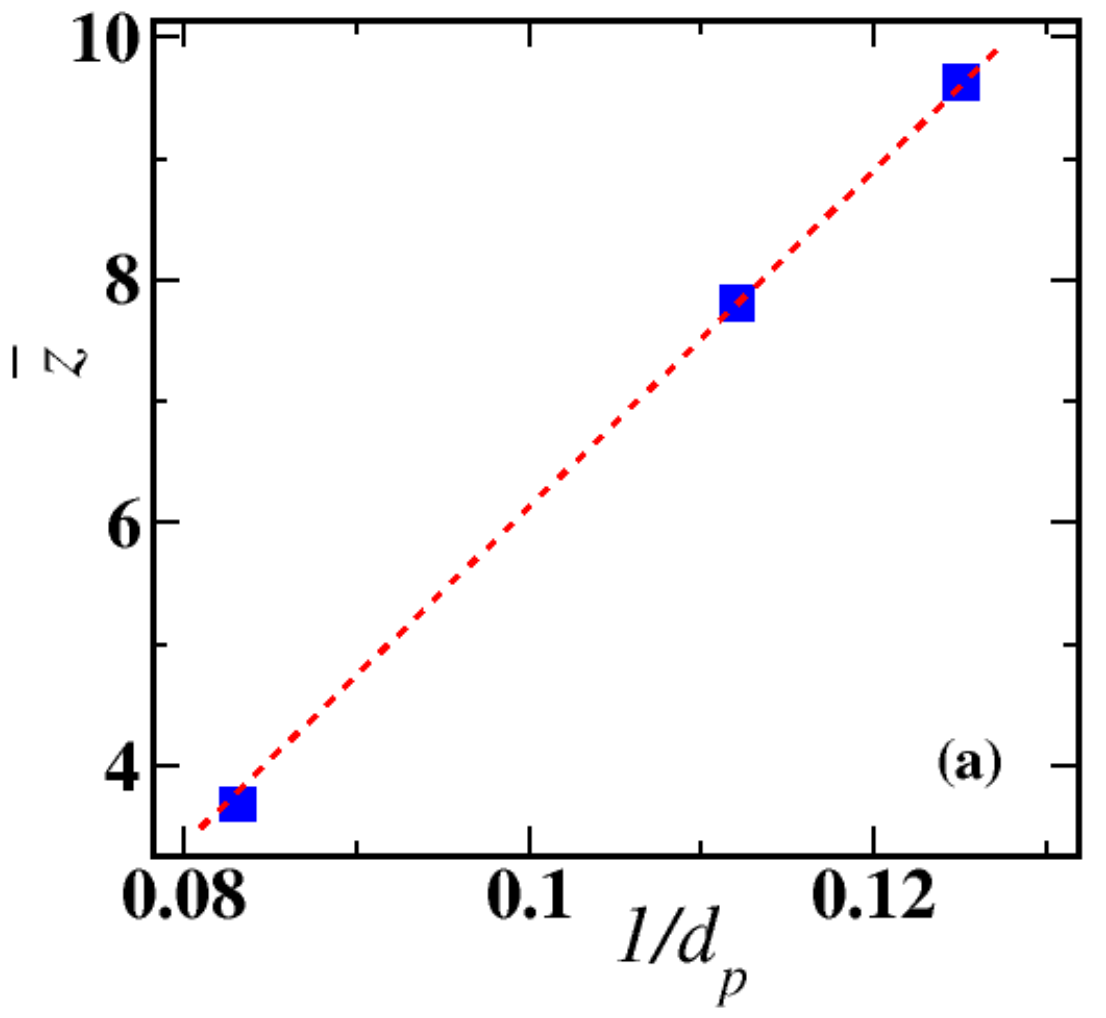}~~~
		{\includegraphics[width=28mm,height=25mm]{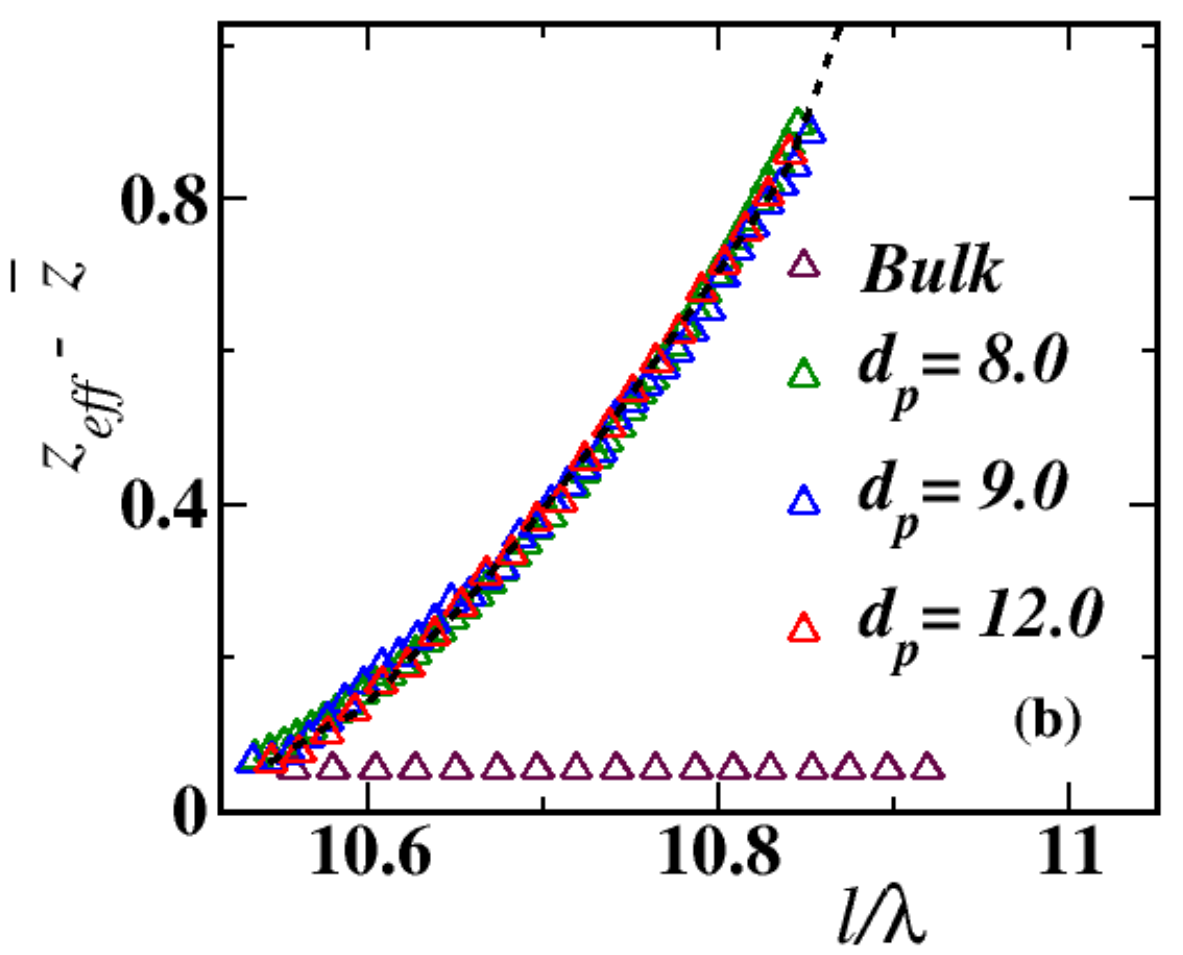}}~
		{\includegraphics[width=28mm,height=25mm]{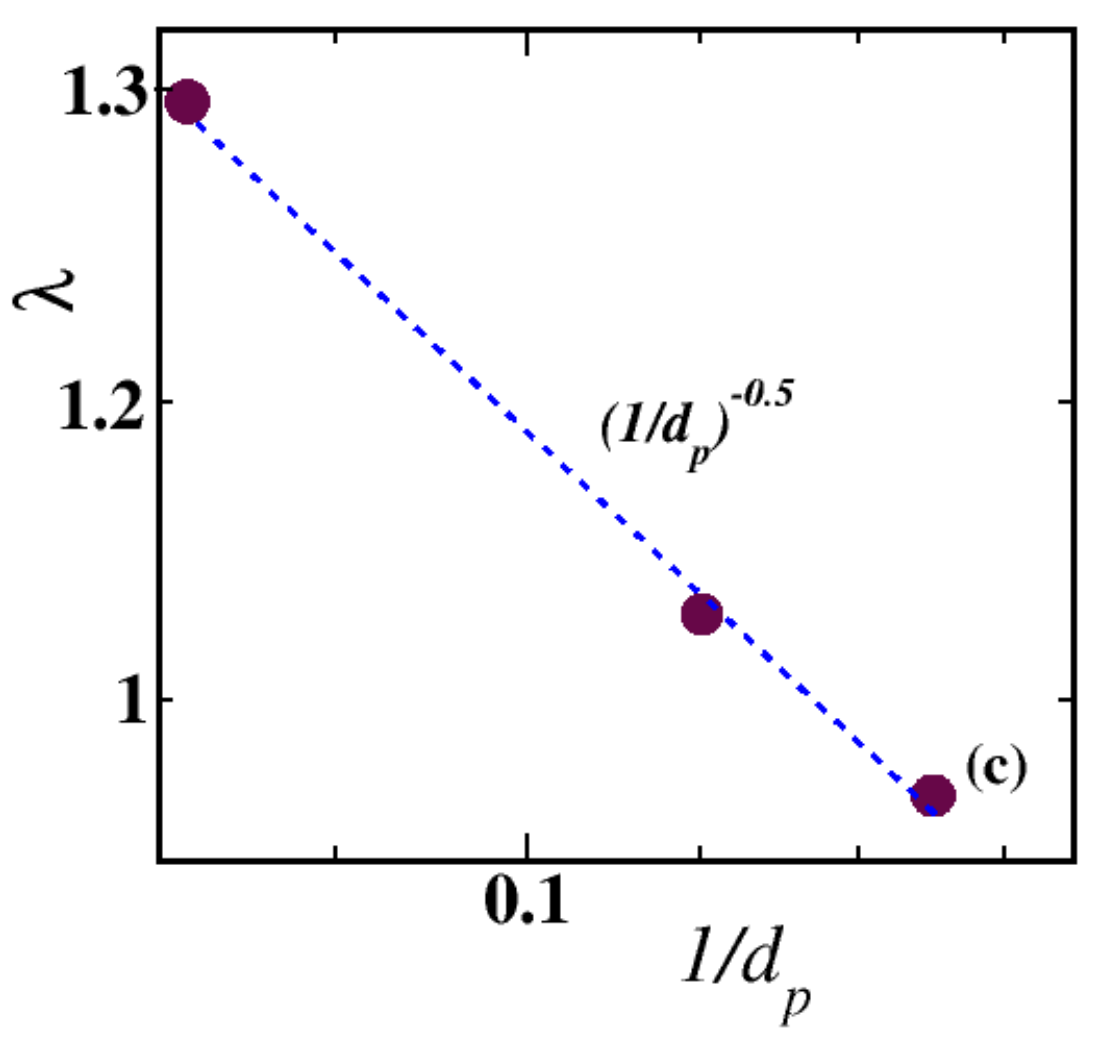}}
	\caption{ (a)  The variation of $\bar{z}$ with $1/d_P$. Dashed line represents the linear best fit curve. (b) ($z_{eff} - \bar{z}$) is plotted as a function of $\ell/\lambda$ for different $d_P$ values. The dashed line represents the best fit curve following Eq.~\ref{scaling}. (c) The variation of $\lambda$ with $1/d_P$ in the log-log scale. Dashed line represents the best fit power law curve with exponent -0.5. }
	\label{fig:5a}
	\label{fig:5b}
	\label{fig:5c}
\end{figure} 

To analyze the deviation from the LS growth law, we apply renormalization-group scaling arguments developed for the random-field Ising model (RFIM)~\cite{CorberiSU1,CorberiSU2}. In the presence of confinement, considering an additional dependence on the pore size, the growth law generalizes to
\vspace{0.001mm}
\begin{equation}
	\ell(t,{d_P}) = t^{\alpha}P(x), ~~~~~~~~~~ x = \frac{d_P}{t^\nu},
	\label{5}
\end{equation}
where $P(x)$ is the scaling parameter, $\alpha = 1/3$ (bulk diffusive exponent) and $\nu$ is a crossover exponent. Eq.~\ref{5} captures the transition from the bulk power-law growth to the asymptotic behavior dictated by confinement, with the nature of this crossover controlled by $\nu$. Recasting the relation yields
\begin{equation}
	\begin{aligned}
		t &= \ell^{1/\alpha} q(y), ~~~~~  q(y)=P^{-1/\alpha}(x), \\
		y &= \frac{\ell}{\lambda},~~~~~~~
	   \lambda = (d_P)^{\alpha/\nu}.
	\end{aligned}
	\label{6}
\end{equation}
The effective growth exponent
\begin{equation}
	z_{\text{eff}}(y)= \frac{\partial~ \mathrm{ln}(t)}{\partial~ \mathrm{ln}(\ell)} = 1/\alpha + \frac{\partial~ \mathrm{ln}~q(y)}{\partial~\mathrm{ln}(y)}.
\end{equation}
identifies whether the dynamics is bulk-like ($\nu>0$) or confinement-dominated ($\nu<0$). Therefore, we compute the instantaneous slope of the length scale curves. As seen in Fig.~\ref{fig:4b}(b), $z_{\mathrm{eff}}$ decreases initially and eventually stabilizes at a constant value in the long-time limit for the bulk diffusive growth. In contrast, confined systems display an intermediate plateau before transitioning to a slower regime. The plateau value $\bar{z}$ increases linearly with $1/d_P$ (Fig.~\ref{fig:5a}(a)), confirming that stronger confinement corresponds to a higher energetic barrier for coarsening~\cite{Paul1,Paul2}.

The crossover regime where the length scale saturates asymptotically is further analyzed by using the following scaling analysis. In Fig.~\ref{fig:5b}(b) we plot the $z_{\mathrm{eff}} - \bar{z}$ against the scaled domain size $\ell/\lambda$. Here $\lambda$ a fitting parameter that depends on the pore size $d_P$. Conspicuously, an excellent data collapse is achieved across pore sizes. The variation of $\lambda$ with $1/d_P$ in Fig.~\ref{fig:5c}(c) reveals the relation $\lambda \sim (1/d_P)^{-0.5}$.  The negative exponent signifies a transition to logarithmic behavior~\cite{CorberiSU1,CorberiSU2}. Thus, the confinement of the porous medium alters the coarsening process, leading to a clear crossover from power-law to logarithmic growth.

To rationalize the scaling behavior observed in Fig.~\ref{fig:5b}(b), we model the effective dynamical exponent using the power-law form~\cite{Ahmad}
\begin{equation}\label{scaling}
	z_{\text{eff}} = \frac{\partial~\mathrm{ln} (t)}{\partial~\mathrm{ln} (\ell)} = \bar{z} + a {\left(\frac{\ell-\ell_0}{\lambda}\right)^\phi}
\end{equation}
where $a \simeq 0.041, \ell_0 \simeq 5.98$ and $\phi \simeq 1.3$ (logarithmic growth exponent) are the fitting parameters. Eq.~\ref{scaling} directly leads to the logarithmic coarsening form 
\begin{equation}\label{15}
\frac{\ell}{\lambda} \sim (\frac{\phi}{b} \mathrm{ln} t)^{1/\phi},~~~~~~~b=0.3. 
\end{equation}
capturing the asymptotic slowdown induced by confinement. The excellent agreement between this scaling function and the numerical data (dashed curve in Fig.~\ref{fig:5b}(b)) provides analytical support for the crossover from a power-law regime to logarithmic growth.

Taken together, these results demonstrate that confinement fundamentally alters the passive phase separation dynamics: pore geometry imposes kinetic barriers that drive a crossover from diffusive power-law growth to a logarithmic regime, ultimately halting coarsening at a pore size restricted length scale. These findings establish the fundamental constraints imposed by geometric confinement on passive phase separation, thereby providing a baseline against which we next evaluate how activity modifies, and in certain regimes overcomes, these limitations in active fluids.
\subsection{ Active Case}
Having established the constraints imposed by geometric confinement on passive phase separation, we now examine how self-propulsion modifies the kinetics and morphology of coarsening in the same porous environment. Here, we focus on a host structure with average pore size $d_P = 12$, and explore a range of activity strengths $f_A$. While representative results are shown for this particular confinement, simulations performed using other pore sizes exhibit consistent behavior, indicating that the influence of activity is robust to variations in geometry.

\begin{figure}[htbp]
	\centering
	\includegraphics[width=40mm,height=35mm]{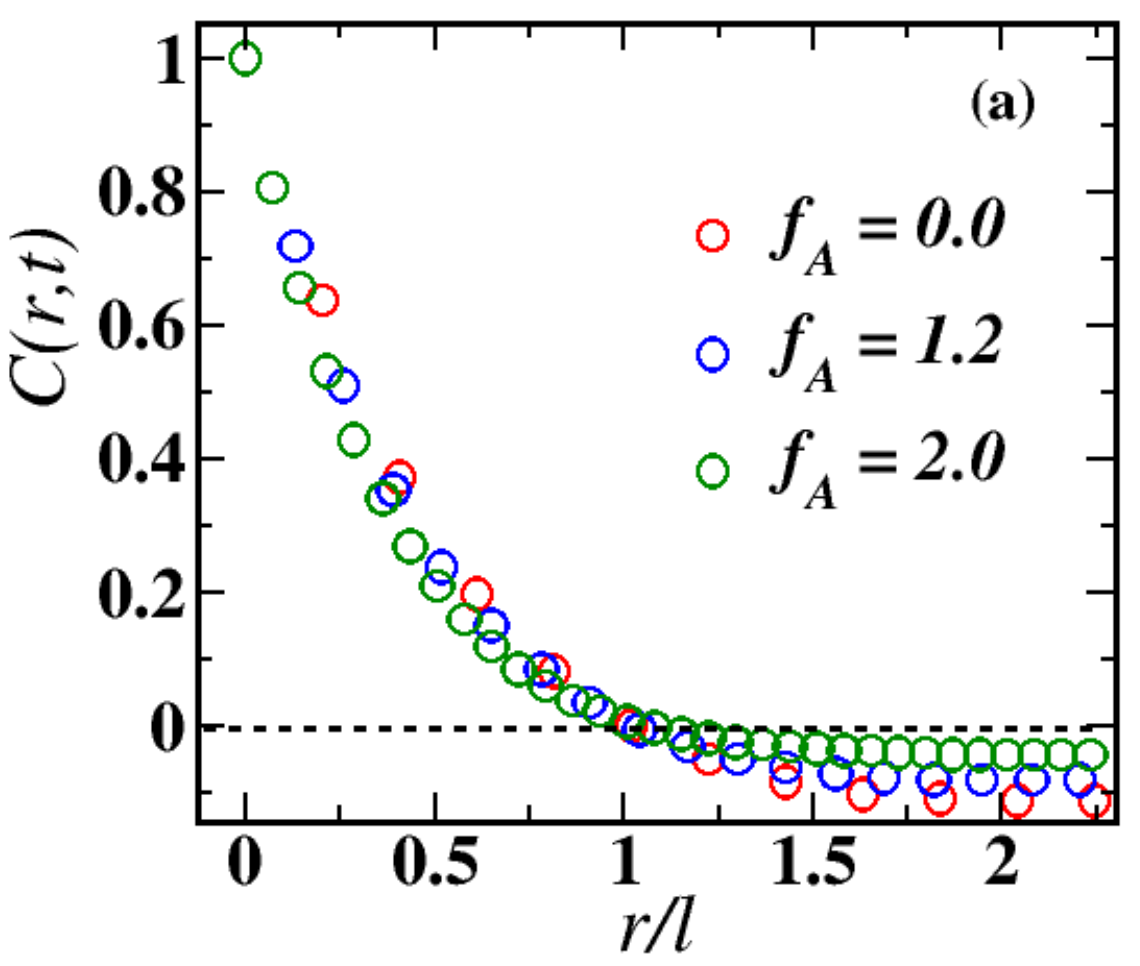}~~~
	\includegraphics[width=40mm,height=35mm]{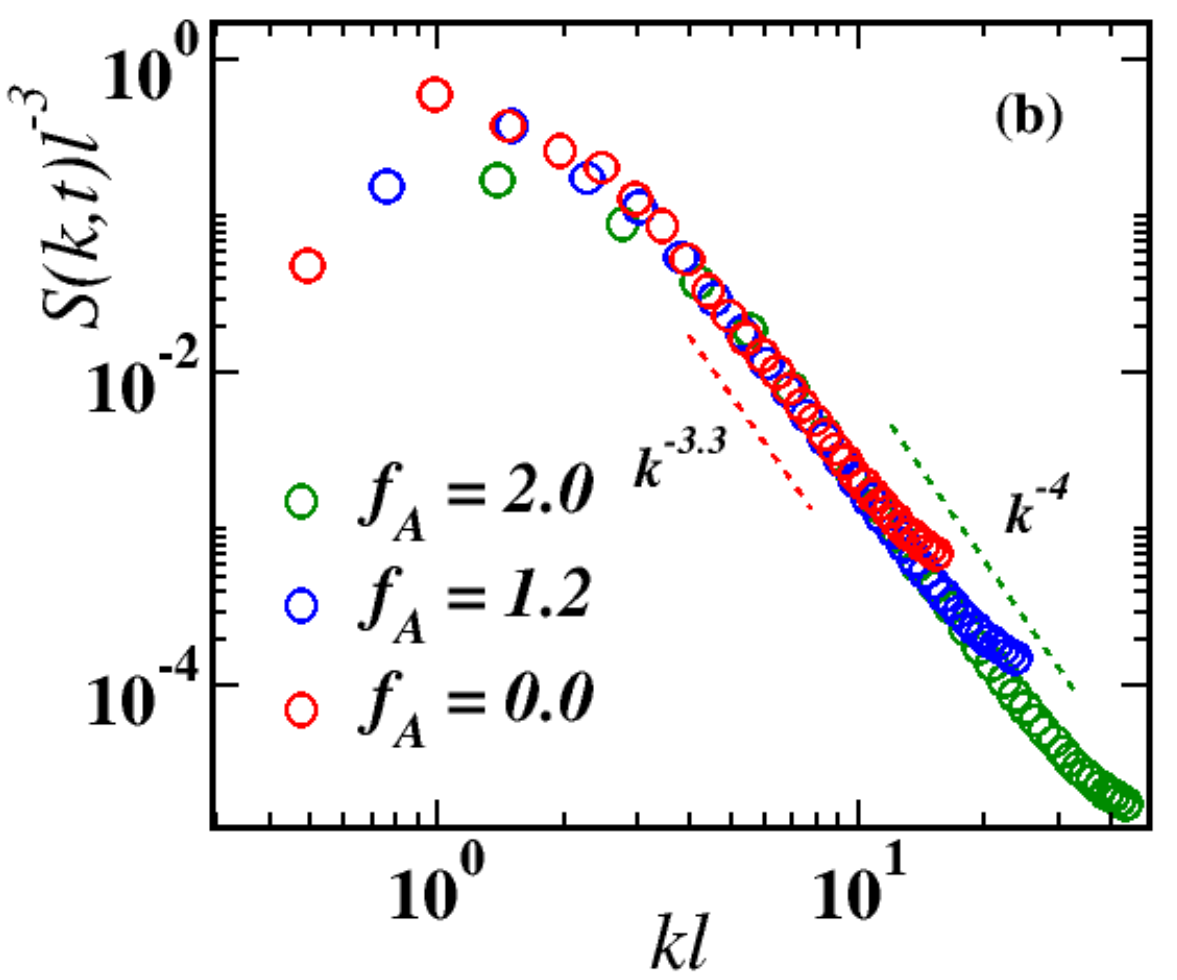}
	\caption{(a) A scaling plot of $C(r,t)$ vs $r/\ell$ for different activity. (b) The scaled structure factor $S(k,t)\ell^{-3}$  vs $k\ell$ for the system in (a) at a fixed time t = 30.}
	\label{fig:6a}
\end{figure}

Figure~\ref{fig:6a} displays the scaled correlation functions and structure factors~\cite{Parameshwaran2} for different $f_A$. In contrast to the passive case, the scaling properties of $C(r,t)$ now depend strongly on activity, demonstrating a clear breakdown of superuniversality with respect to the activity field. Moreover, the structure factor reveals that increasing activity progressively restores Porod-like behavior ($S(k,t)\sim k^{-4}$), suggesting that active stresses sharpen the interfaces and suppress the fractal roughening induced by confinement~\cite{Gaurav}. This trend implies that motility mediates more efficient rearrangements of constituent particles, facilitating the growth of smoother domain boundaries.

\begin{figure}[t]
	\centering
	\includegraphics[width=28mm,height=28mm]{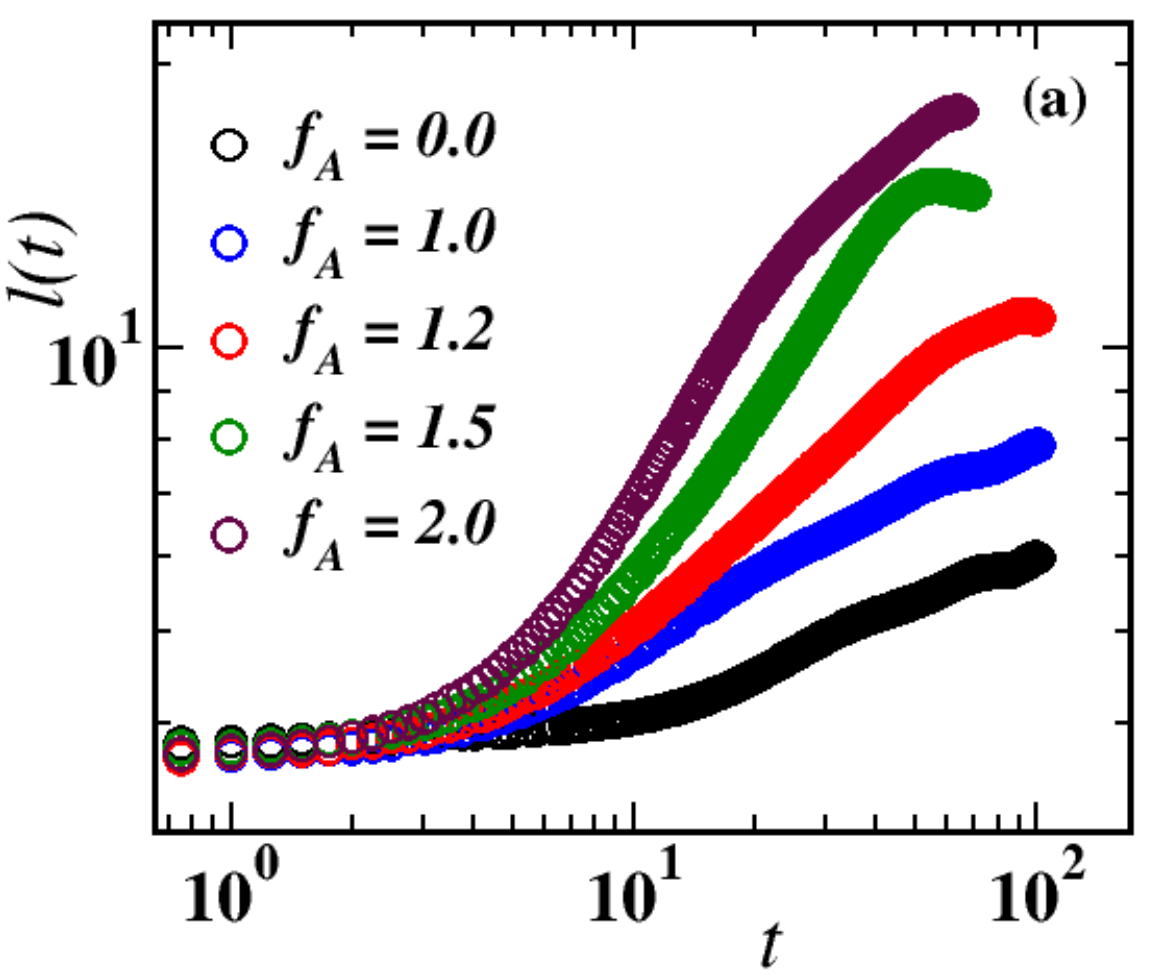}~
	\includegraphics[width=30mm,height=28mm]{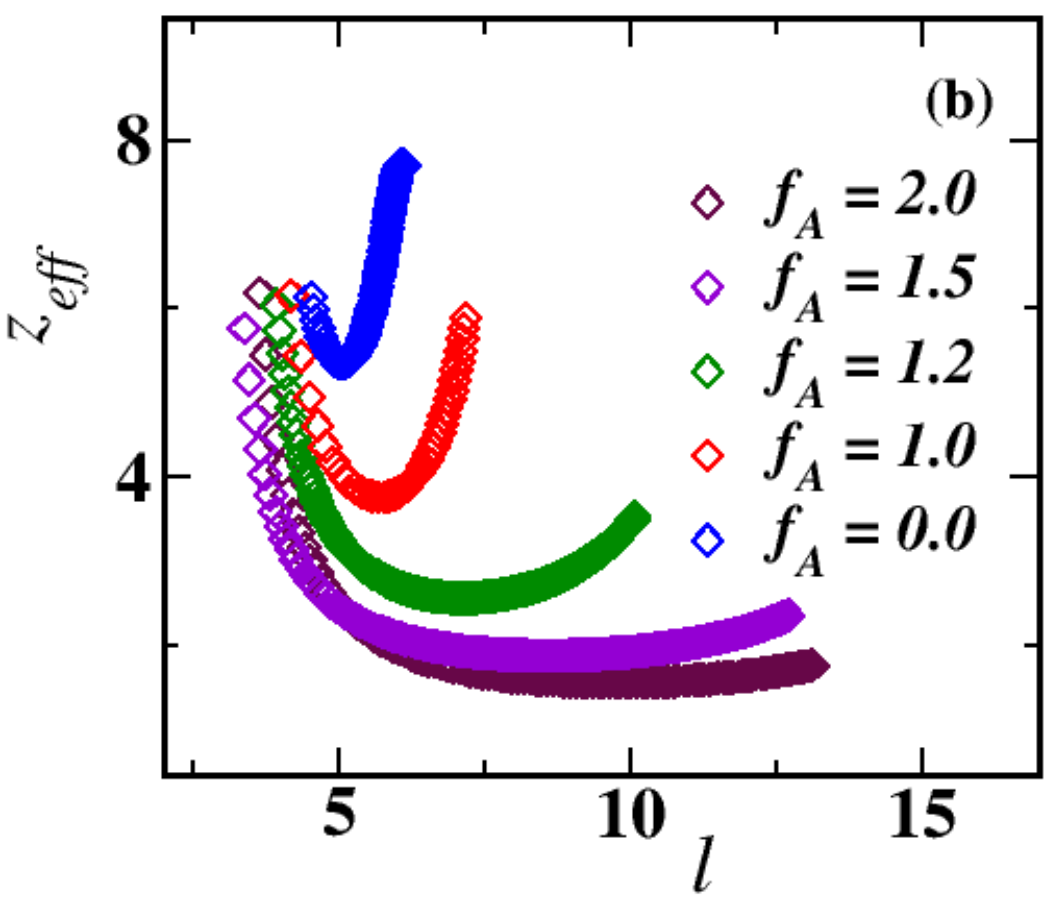}~
	\includegraphics[width=28mm,height=28mm]{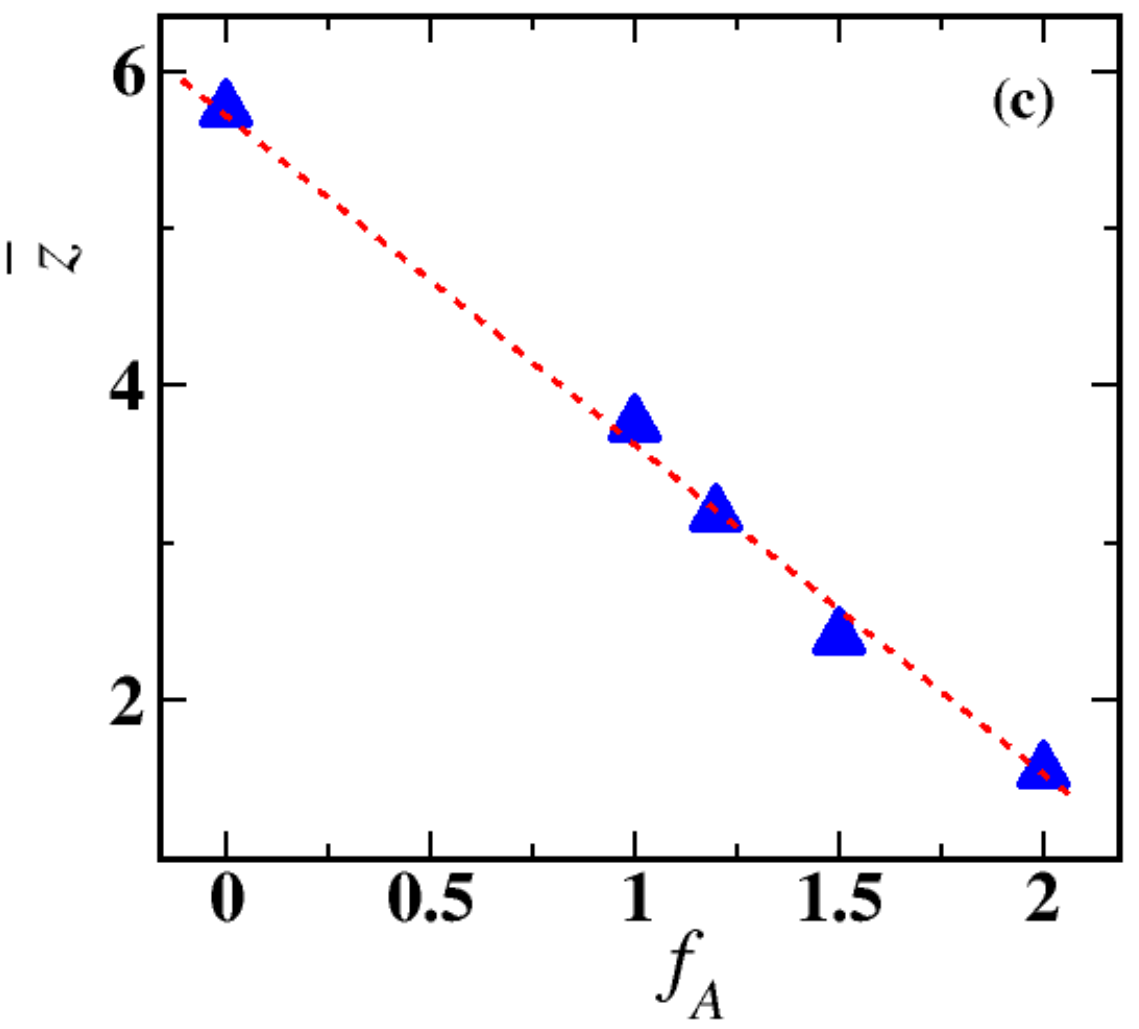}
	\caption{(a) The length scale $\ell(t)$ is shown on a log-log plot for various activity strengths over time. (b) Plot of instantaneous dynamical exponent $z_{\mathrm{eff}}$ vs. $\ell$ for different activity. (c) Plot of $\bar{z}$ vs. $f_A$. Dashed line represents the linear best fit curve.}
	\label{fig:7}
\end{figure}

The resulting enhancement in coarsening kinetics is evident in Fig.~\ref{fig:7}(a), where $\ell(t)$ increases more rapidly with increasing $f_A$. Notably, for $f_A = 2$, the growth exponent approaches unity, corresponding to hydrodynamic like growth. Thus, activity reinstates fast coarsening dynamics even in environments where passive systems become kinetically arrested at long times.

To quantify this crossover, we analyze the instantaneous growth exponent $z_{\text{eff}}$, as in the passive case. Figure~\ref{fig:7}(b) shows that low-activity systems exhibit a time-dependent crossover similar to the confined passive fluid. However, beyond a threshold $f_A$, the exponent saturates to a constant value, signaling persistent power-law growth with no indication of late-time arrest. The dependence of the intermediate-regime exponent on $f_A$ is displayed in Fig.~\ref{fig:7}(c), revealing a linear change, highlighting the direct and tunable role of activity in accelerating coarsening.

\begin{figure}[t]
	\centering
      \raisebox{1.2mm}{\includegraphics[width=40mm,height=36mm]{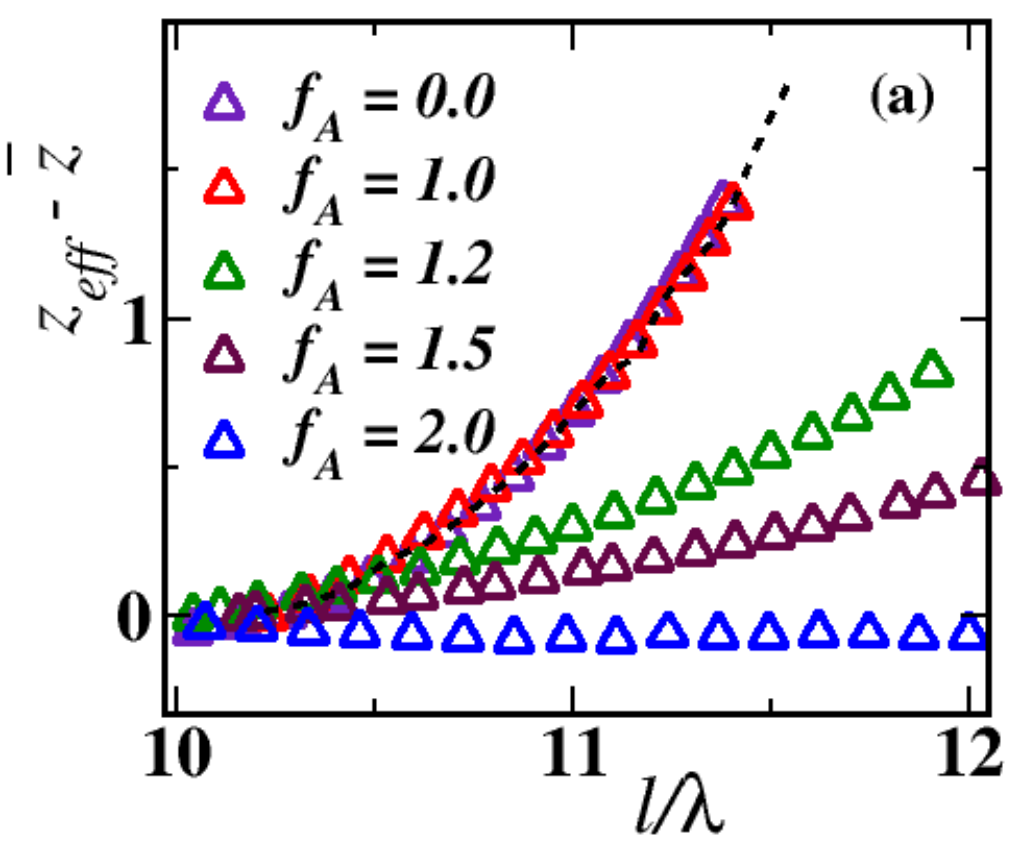}}~~
      \includegraphics[width=40mm,height=37mm]{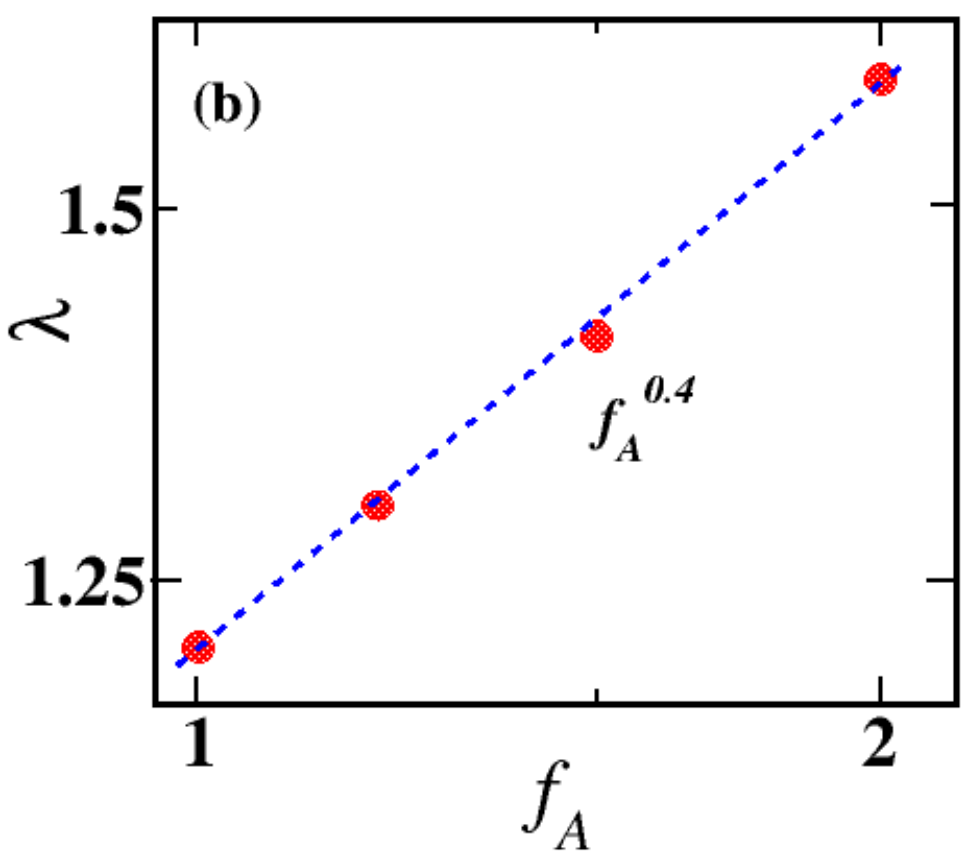}
	\caption{(a) ($z_{eff} - \bar{z}$) plotted as a function of scaled length $\ell/\lambda$ for different $f_A$. (b) The variation of $\lambda$ with $f_A$ in the log-log scale. The dashed line represents the best fit curve.}
	\label{fig:8}
\end{figure}

We further explore the long-time regime by rescaling the data using the pore-dependent scaling length $\lambda$ derived for the passive system. As shown in Fig.~\ref{fig:8}(a), systems with $f_A \leq 1$ collapse cleanly onto a universal curve consistent with logarithmic coarsening, in agreement with the analytical form given by Eq.~\ref{scaling} with fitting parameters
$a\simeq 0.11$, $\ell_0\simeq 6.22$, $\phi\simeq 1.5$, $b\simeq 0.1$. At higher activities, however, the curves depart significantly from this scaling form. For $f_A=2$, $z_{\text{eff}}$ remains constant at late times, demonstrating complete suppression of logarithmic slowdown and full recovery of ballistic coarsening despite confinement. The deviations exhibit a power-law behavior, as verified in Fig.~\ref{fig:8}(b), where the data show a linear increase with activity with an exponent 0.4. The positive exponent indicates that the system approaches bulk-like growth, consistent with Eq.~\ref{6}. This confirms that active driving systematically overcomes pore-induced kinetic barriers and shifts the dynamics toward faster, power-law growth.

Taken together, these results reveal that activity plays a dual role in confined phase separation: it restores smooth interfacial morphology otherwise disrupted by porous complexity, and crucially it re-enables rapid coarsening that is fundamentally inaccessible to passive systems under the same geometric constraints. Thus, confinement no longer dictates the kinetic fate of the system once self-propulsion becomes sufficiently strong, pointing to an activity-controlled pathway for unlocking phase separation within structurally heterogeneous environments.

\section{Conclusion and Outlook}
In this work, we have provided a comprehensive investigation of vapor-liquid phase separation inside complex porous media for both passive and active fluids. By systematically tuning the porous architecture through a freeze-quench protocol, we revealed how confinement generates kinetic bottlenecks in passive systems, leading to dramatically slowed coarsening and ultimate morphological arrest. Using scaling and renormalization arguments, we showed that diffusive growth transitions to a logarithmic regime at long times, with domain sizes saturating at pore-size–restricted limits. These results establish the intrinsic limitations imposed by geometric complexity on passive phase separation dynamics.

Introducing self-propulsion fundamentally alters this paradigm. Activity not only accelerates coarsening but also restores sharp interfaces and suppresses the roughened, fractal morphologies characteristic of confined passive systems. At sufficiently high activity, logarithmic slowdown is completely removed and ballistic growth re-emerges, despite the same geometric constraints. Self-propelled particles effectively “unlock’’ confined domains, enabling rapid restructuring of dense and dilute regions and overcoming the kinetic barriers imposed by pore topology.

These findings highlight the profound and tunable role of activity in determining the fate of phase separation in heterogeneous environments. They bridge concepts from nonequilibrium physics, soft condensed matter, and materials science, and have important implications for a wide range of systems in nature and technology. For instance, vapor-liquid coexistence in porous geological materials, droplet organization in emulsions and foams, nutrient or chemical segregation in biological tissues, and the behavior of active colloids and microorganisms in crowded microenvironments all rely on similar underlying mechanisms. Our results suggest that biological and synthetic microswimmers may exploit activity to reorganize fluids or create spatial patterns in confined spaces, an essential feature in cellular phase separation, biofilm architecture, or active transport in the extracellular matrix.

More broadly, this study demonstrates how the interplay of geometry and activity can govern the emergence of organization in disordered environments. Extending these concepts to anisotropic, dynamic, or hierarchically structured porous networks could yield new design principles for smart materials, targeted delivery strategies, or microfluidic devices that harness or regulate active forces. Future opportunities include incorporating chemical signaling, viscoelasticity of the host matrix, and feedback loops between confinement and activity, key elements in living systems.
Overall, our results establish a general physical framework for understanding and controlling phase separation in complex environments, offering a route to engineer nonequilibrium organization in both biological and artificial active systems.

\textit{Acknowledgements.}~~B.S.G. acknowledges the Science and Engineering Research Board (SERB), Department of Science and Technology (DST), Government of India (No. CRG/2022/009343) for financial support. P.M. acknowledges the support by the INSPIRE Fellowship (No. DST/INSPIRE Fellowship/2023/IF230398) of the Department of Science and Technology (DST), Government of India.

\end{document}